\documentclass[10pt,twocolumn]{article} 

\pdfoutput=1

\usepackage{latex8,float}
\usepackage{times,amsmath,subfigure}
\usepackage{url,algorithmic,algorithm,graphicx}
\usepackage[T1]{fontenc}
\usepackage[latin1]{inputenc} 
\usepackage[small,nooneline,bf,hang]{caption}
\usepackage{stfloats}

\begin{document} 
\title{Distributed Creation and Adaptation of Random Scale-Free Overlay Networks}

\author{Ingo Scholtes\\
University of Trier\\ Systemsoftware and Distributed Systems \\ D-54286 Trier, Germany\\ scholtes@syssoft.uni-trier.de\\http://syssoft.uni-trier.de/scholtes \\
}

\maketitle
\thispagestyle{empty}

\addtolength{\abovedisplayskip}{-6mm}
\addtolength{\belowdisplayskip}{-1mm}
\addtolength{\abovecaptionskip}{-3mm}
\addtolength{\belowcaptionskip}{-3mm}
\addtolength{\dblfloatsep}{-7mm}
\addtolength{\intextsep}{-1mm}
\addtolength{\textfloatsep}{-2mm}
\addtolength{\dbltextfloatsep}{-2mm}
\setlength{\tabcolsep}{3pt}

\linespread{0.89}

\begin{abstract}
Random scale-free overlay topologies provide a number of properties like for example high resilience against failures of random nodes, small (average) diameter as well as good expansion and congestion characteristics that make them interesting for the use in large-scale distributed systems. A number of these properties have been shown to be influenced by the exponent $\gamma$ of their degree distribution $P(k)\propto k^{-\gamma}$. In this article, we present a distributed rewiring scheme that is suitable to effectuate scale-free overlay topologies with an adjustable exponent. The scheme uses a biased random walk strategy to sample new endpoints of edges being rewired and relies on the equilibrium model for scale-free networks presented in \cite{Goh2001}. The bias of the random walk strategy can be tuned to produce random scale-free networks with arbitrary degree distribution exponents greater than two. We argue that the rewiring strategy can be implemented in a distributed fashion based on a node's information about its immediate neighbors. We present both analytical arguments as well as results that have been obtained using an implementation of the proposed protocol.
\end{abstract}

\Section{Motivation}

During the last decade, the increasing spread and importance of large-scale Peer-to-Peer systems has raised significant research interest in the design and analysis of robust and efficient overlay networks. In this research, structured and unstructured approaches can be distinguished. Mimicking the use of data structures in traditional computing, highly structured overlay topologies facilitate the use of efficient distributed algorithms with deterministic performance. However, the overhead entailed by the construction and maintenance of such deterministically structured topologies questions their usability in large-scale scenarios with dynamic and potentially faulty participants. Constituting a different approach, unstructured overlay networks do not impose constraints about the detailed structure of the emerging network topology. Rather than using costly and potentially complex routines for building and maintaining sophisticated network structures, in such unstructured overlays links can arise in a seemingly random and uncoordinated fashion. They are thus particularly suitable for highly dynamic scenarios in which the operational overhead entailed by structured approaches can possibly dominate a system's overall performance \cite{Balakrishnan2003}.

While the use of unstructured overlays can reduce construction and maintenance overhead, designing efficient distributed algorithms with predictable performance is hardly possible when making no assumptions whatsoever about an overlay's structure. Interestingly, based on a stochastic model of the system in question and arguments from random graph theory and complex network science, it is often possible to reason about structural properties of the resulting network topology that hold \emph{almost surely} in the limit of large systems. Similarly, the performance of a number of dynamical processes - many of them relevant to distributed computing systems - has been studied in random network structures. For sufficiently large systems, based on randomized overlay topologies one can thus obtain strong, though probabilistic guarantees about their structure and performance. Considering the classical taxonomy of deterministically structured and completely unstructured overlay networks, this suggests an intermediate class of \emph{probabilistically structured} topologies that promises to combine the benefits of both.

During the last decade, much of the work in the field of random networks has been focused around scale-free networks that are characterized by a power law degree distribution $P(k)\propto k^{-\gamma}$. The fact that networks with such scale-free characteristics seem to emerge naturally in a variety of natural, social and technological contexts has awakened the interest of researchers in disciplines as diverse as mathematics, statistical physics, biology, sociology, and computing. It has since been shown that scale-free networks provide a number of interesting properties like a remarkable robustness against random failures \cite{Albert2000,Dorogovtsev2000}, small diameter and average path lengths \cite{Cohen2003a,Bollobas2004} as well as favorable expansion and congestion properties \cite{Gkantsidis2003,Toroczkai2004}. Some of these properties make them interesting for the design of large-scale computing systems and - in fact - for certain networked computing systems it has been observed that scale-free structures emerge in a seemingly self-organized way \cite{Albert1999,Faloutsos1999,Medina2000,Ripeanu2002}. 

Based on this observation, during the last couple of years, the performance of distributed algorithms operating in scale-free networks has been studied. For the problems of distributed search \cite{Adamic2001,Lv2002,Boykin2004}, information dissemination and entropy reduction protocols \cite{Jelasity2005} as well as synchronization \cite{Scholtes2010}, distributed schemes have been derived that seem to work particularly efficiently in scale-free networks. Considering a scale-free network topology with a degree distribution $P(k)^{-\gamma}$, it has further been argued that the exponent $\gamma$ has massive influence on network properties like diameter \cite{Cohen2003a}, the vulnerability against targeted attacks as well as the performance of dynamical processes \cite{Barrat2008}. The reason for this can be found in the fact that the exponent $\gamma$ determines the skewness of the degree distribution and thus the frequency and magnitude of highly connected hub nodes. For the practical design of scale-free overlay topologies, the degree distribution exponent is thus a critical parameter which largely influences their robustness, the performance of distributed algorithms as well as the distribution of load being imposed on individual machines.

In this article, we study how systems with scale-free overlay structures can adapt the degree distribution exponent and thus tune the heterogeneity of overlay connectivity in a distributed and directed fashion while maintaining the overall scale-free characteristics of their topology. Based on analogies to equilibrium and non-equilibrium statistical physics that have been put forth in the study of complex networks\cite{Albert2002,Farkas2004} and the fact that - in the limit of large systems - a number of network properties change abruptly when the degree distribution exponent exceeds certain critical points, one may view such a mechanism as an active triggering of network phase transitions \cite{Scholtes2008a}. In this article, we discuss a simple distributed rewiring mechanism that can be used for this purpose in non-growing overlay topologies. It is based on the equilibrium model of uncorrelated scale-free networks that has been considered in \cite{Goh2001,Lee2004a} and makes use of a biased random walk strategy in order to sample the endpoints of edges being rewired. As we shall see later, the efficiency and thus feasibility of the mechanism is based on the favorable expansion properties of certain classes of random networks. A detailed description and derivation of the proposed protocol will be presented in section \ref{sec:RandomWalk}. Here we further present some analytical arguments on the convergence behavior of the random walk sampling strategy underlying the protocol being presented in the subsequent section \ref{sec:Protocol}. In section \ref{sec:Evaluation} we present simulation results that have been obtained using an implementation of the proposed rewiring scheme. Having briefly reviewed related work, in section \ref{sec:Conclusion} we conclude the article by summarizing its contribution and pointing out a number of open issues and threats to validity.

\Section{Creating and Adapting Scale-Free Overlays}
\label{sec:RandomWalk}

As has been argued above, the exponent $\gamma$ is a macroscopic, statistical parameter that influences the structural properties of random networks with a power law degree distribution $P(k) \propto k^{-\gamma}$. In the following, we thus intend to derive a distributed protocol that can be used to effectuate scale-free network topologies with a particular degree distribution exponent. As initial situation, we assume an arbitrary, connected overlay topology. While for the functioning of the scheme no particular initial state of the overlay is required as long as it is connected, it will later be argued that the initial topology influences the efficiency of the scheme in terms of the number of messages that need to be exchanged. For simplicity, we further assume that each of the $n$ nodes is uniquely identified by a numeric identifier $i \in \{1, \ldots, n\}$. However in sufficiently large systems, per-node quantities $i$ that are chosen uniformly at random - and thus not necessarily unique - can be used instead. In order to simplify derivation and analysis, we further consider a static situation in which no nodes enter or leave the system. Clearly, the main motivation to use a probabilistic overlay topology in the first place is to support highly dynamic systems in which node joins and exits are frequent. Although in this article we only consider the simpler static situation, we argue that our results can readily be extended to dynamic situations with fluctuating participants.

In order to derive a distributed scheme that can be used to influence the structure of scale-free overlays, we first need a model that is capable of generating network topologies with tunable power law degree distribution exponents. Here we use a simple equilibrium model for scale-free networks with a fixed number of nodes that has been introduced in \cite{Goh2001} and analyzed in \cite{Lee2004a}. In this model it is assumed that each node $i \in \{1, \ldots, n\}$ is assigned a weight $w_i = i^{-\alpha}$ for some parameter  $\alpha$ in the range $(0,1)$. It is then assumed that $m$ edges are created between pairs of nodes $(i,j)$ chosen with probabilities $p_i$ and $p_j$ that are given by the normalized weights

\begin{equation}
\label{equ:prob}
p_i=\frac{w_i}{\sum_{k=1}^{n}{w_k}}.
\end{equation}

As has been argued in \cite{Lee2004a}, this simple model produces uncorrelated random scale-free networks with a degree distribution 

\begin{equation*}
	P(k) \propto k^{- (1 + \frac{1}{\alpha})}
\end{equation*}

Hence, for $\alpha \rightarrow 0$, the model yields a scale-free network with the exponent $\gamma \rightarrow \infty$, while for $\alpha \rightarrow 1$ the exponent $\gamma$ converges to two. Hence, it provides a parameter that can be adjusted to effectuate scale-free networks with an arbitrary degree distribution exponent $\gamma$ in the range $(2,\infty)$.

In order to apply this simple model in practical networked systems, a distributed scheme is required that creates edges between two nodes $i$ and $j$ in an overlay network with probability $p_ip_j$. For this, we assume that we start with a random, connected overlay topology consisting of $n$ nodes and $m$ edges. In practice, this initial topology may emerge by means of an arbitrary bootstrapping method that connects joining nodes to existing participants either deterministically or at random. We can then view the above model as a rewiring scheme that gradually replaces existing edges so that edges between node pairs emerge with the desired node-dependent probabilities. For this, a node initiating the rewiring of an edge must be able to sample two new endpoints for the edge being rewired according to the probability measure given in equation \ref{equ:prob}. While one can imagine different mechanisms by which this can be achieved \cite{Jelasity2007}, a simple and promising method to sample nodes in Peer-to-Peer systems is by means of a random walk \cite{Zhong2005,Gkantsidis2006}. For this we consider that nodes wishing to rewire an edge sample two new endpoints by means of two independent random walks through the current network topology. For a classical, unbiased random walk, the probability $\pi_i(t)$ to find the walker at an arbitrary time $t$ at node $i$ converges to

\begin{equation*}
	\pi_i(t) \rightarrow \frac{d_i}{N \cdot \bar{d}} \text{  }(t \rightarrow \infty)
\end{equation*}

where $\bar{d}$ is the average node degree of the network. In order to sample nodes with the probabilities given in equation \ref{equ:prob} we need to introduce a random walk bias that influences the transition probabilities accordingly. Considering a random walk in a connected overlay topology $G(V,E)$ as Markov chain with state space $V$ and stationary distribution $\pi$, the random walk bias can be configured by means of a Metropolis-Hastings chain \cite{Metropolis1953,Hastings1970,Azar1992} in such a way that a desired stationary distribution $\pi$ holds. In general, this can be achieved by introducing a bias as shown in the following transition matrix $T$:

\begin{equation}
\label{equ:Metropolis}
T_{ij} = 
\begin{cases}
	\frac{1}{d_i} \text{min}\left\{ \frac{\pi_j}{\pi_i} \frac{d_i}{d_j}, 1 \right\} & (i,j) \in E, i\ne j \\
	1 - \frac{1}{d_i} \sum_{(k,i)\in E}{P_{ik}} & i = j \\
	0 & (i,j) \notin E \\
\end{cases}
\end{equation}

Here $d_i$ denotes the current degree of node $i \in V$ and an entry $T_{ij}$ gives the probability that a random walk residing at node $i$ moves to node $j$. The fact that this transition matrix has stationary distribution $\pi$ follows from the reversibility of the underlying Markov chain, as well as from its irreducibility (assuming a connected network topology) and aperiodicity (self-loops are possible). Under these restrictions, the Markov chain convergence theorem ensures that the probability $\pi_i(l)$ to find a random walker that has been started in an arbitrary node resides at node $i$ after $l$ steps converges to $\pi$ as $l$ goes to infinity.

From this, one can easily configure a random walk bias that results in a stationary distribution suitable to sample nodes in a way that - after rewiring - a scale-free network with degree distribution exponent $\gamma$ emerges. From the probability $p_i$ in equation \ref{equ:prob} and the fact that it gives rise to a scale-free network with degree distribution exponent $1+\frac{1}{\alpha}$, we obtain the desired stationary distribution

\begin{equation}
\pi_i^\gamma = \frac{i^\frac{-1}{\gamma-1}}{\sum_{k=1}^{n}{k^\frac{-1}{\gamma-1}}}
\end{equation}

which, with equation \ref{equ:Metropolis} and $\frac{\pi_j}{\pi_i} = \left( \frac{i}{j} \right)^{\frac{1}{\gamma -1}}$, yields the following transition matrix $P$:

\begin{equation}
\label{equ:bias}
P_{ij} = 
\begin{cases}
	\frac{1}{d_i} \text{min}\left\{ \left( \frac{i}{j}\right)^{\frac{1}{\gamma-1}} \frac{d_i}{d_j}, 1 \right\} & (i,j) \in E, i\ne j \\
	1 - \frac{1}{d_i} \sum_{(k,i)\in E}{P_{ik}} & i = j \\
	0 & (i,j) \notin E \\
\end{cases}
\end{equation}

Thus, a random walk with the above bias can be used to sample endpoints of edges and thus perform rewiring operations that effectuate scale-free network topologies with a particular degree distribution exponent.

\SubSection{Bounding the Random Walk Length}
\label{sec:Bound}

The goal of this article is to practically apply the above strategy in a distributed rewiring scheme. Hence, an important question that needs to be answered is how many steps a random walk with the above bias needs to take before the probability $\pi_i(l)$ to find it in a node $i$ after $l$ steps is sufficiently close to the desired stationary limit $\pi_i$. In the rewiring protocol presented in the following section, this translates to the number of messages that need to be exchanged for a single rewiring operation. To assess this convergence behavior, one first needs to give a formal definition of when two probability distributions $\pi$ and $\pi'$ shall be considered \emph{sufficiently close}. For this we use the usual definition of the \emph{total variation distance} $D$ which - for two probability measures $\pi$ and $\pi'$ and a finite state space $V$ - is defined as follows:

\begin{equation*}
D(\pi',\pi) = \frac{1}{2} \sum_{v \in V}{|\pi'(v) - \pi (v)|}
\end{equation*}

The configuration of the random walk bias according to equation \ref{equ:bias} and the Markov convergence theorem ensure that $D(\pi(l), \pi)\rightarrow 0$ for $l \rightarrow \infty$. For an arbitrarily chosen total variation distance $\epsilon>0$ we can then assess the number of steps $l$ our random walk needs to take until $D(\pi(l),\pi) \leq \epsilon$. In order to bound the minimally required number of steps $l$, the arguments put forth in \cite{Sinclair1992} can be used. Here it is argued that an upper bound for $l$ is given by 

\begin{equation*}
	l \leq \frac{ ln \left(\frac{1}{\pi_s \epsilon} \right)}{1-|\lambda_2(P)|}
\end{equation*}

where $\pi$ is the stationary distribution of the Markov chain, $\lambda_2(P)$ is the second smallest eigenvalue of the transition matrix $P$ and $s$ is the initial state. Thus, finding an upper bound for the number of random walk steps requires to find a lower bound for the second smallest eigenvalue $\lambda(P)$ of the transition matrix. Unfortunately, obtaining good bounds for the eigenvalues of stochastic matrices is a non-trivial task. Nevertheless, based on the canonical path approach introduced in \cite{Diaconis1991,Sinclair1992}, analytical arguments concerning the convergence behavior of random walks with a Zipf stationary distribution have been put forth in \cite{Zhong2005,Zhong2008}. In the following we briefly repeat these arguments for the particular random walk strategy considered in this article. In \cite{Zhong2008} it has been argued that, if the stationary distribution $\pi$ is highly skewed, a lower bound for the eigenvalue gap $1-|\lambda_2(P)|$ is given by

\begin{equation*}
	1-|\lambda_2(P)| \geq \frac{\pi_{min}}{D \cdot d_{max}}.
\end{equation*}

Here $D$ denotes the diameter of the network topology upon which the random walk operates, $\pi_{min}$ is the minimum probability ascribed to any vertex by the stationary distribution and $d_{max}$ is the maximum degree of any vertex in the network. Thus, for the special case of Zipf stationary distributions, an asymptotic upper bound for the random walk length $l$ required to achieve a total variation distance smaller than $\epsilon$ is given as \cite{Sinclair1992,Zhong2008}:

\begin{equation}
\label{equ:boundA}
l \leq  ln \left(\frac{1}{\pi_s \epsilon} \right) \cdot \frac{D \cdot d_{max}}{\pi_{min}}
\end{equation}

For a random walk strategy configured to eventually effectuate a degree distribution exponent $\gamma$ and thus stationary distribution $\pi^\gamma$, for the inverse stationary probability of the starting node $s$, the following bound holds:

\begin{equation*}
\frac{1}{\pi_s^\gamma } = s^\frac{1}{\gamma-1} \cdot \sum_{k=1}^{n}{k^\frac{-1}{\gamma-1}} \leq s^\frac{1}{\gamma-1} \cdot \sum_{k=1}^{n}{1} = n \cdot s^\frac{1}{\gamma-1}
\end{equation*}

While this holds for arbitrary $\gamma \in [2,\infty)$ and starting nodes $s$, for the special case of node $n$ we can give a better bound by observing that - due to the increasing skewness - node $n$ is ascribed minimal probability for $\gamma=2$, that is for $\gamma \in [2,\infty)$

\begin{equation*}
	\pi_{min}^\gamma \geq \pi_{min}^{\gamma=2}
\end{equation*}

holds. With this, we can bound the inverse minimal probability by considering the logarithmic growth of the harmonic series, so that

\begin{equation*}
	\frac{1}{\pi_{min}^\gamma} \leq \frac{1}{\pi_{min}^{\gamma=2}} = n \cdot \sum_{k=1}^{n}{\frac{1}{k}} = n \cdot H_n = n \cdot ( ln(n) + \tau + r_n )
\end{equation*}

where $\tau$ denotes the Euler-Mascheroni constant and $r_n \rightarrow 0$ in the limit of large $n$. Assuming an initial scale-free topology with $n$ nodes and degree distribution exponent $\gamma_i$ allows to asymptotically bound diameter and maximum degree as $O(ln(n))$ and $O(n^\frac{1}{\gamma_i})$ respectively \cite{Zhong2008}. Thus, for large $n$ and a random walk started in node $s$, an asymptotic upper bound for the minimal length $l$ to achieve total variation distance smaller than $\epsilon$ can be given as follows:

\begin{equation}
	\label{equ:boundB}
	l = O\left( ln\left( \frac{n \cdot s^{\frac{1}{\gamma-1}}}{\epsilon}\right) \cdot ln(n)^2 \cdot n^{1+ \frac{1}{\gamma_i}} \right)
\end{equation}

This theoretic bound scales worse than linear with the network size $n$. However, as has previously been observed e.g. in \cite{Zhong2008}, the underlying bounding technique is not necessarily tight, that is the actual convergence behavior of a random walk can be considerably better. Since at present obtaining tight upper bounds for the convergence of Markov chains in complex network topologies is an open research issue, in section \ref{sec:Evaluation} we present simulations that have been performed to derive practicable random walk lengths empirically. As will be argued later, the results of these simulations suggest that the adaptation scheme presented in this article can be practically implemented with reasonable random walk lengths. Although these results suggest that the analytical bounds shown above are not tight and thus uninformative with respect to the performance of the scheme in practice, they can nevertheless be used to study by which parameters the convergence behavior of a random walk is influenced. From equation \ref{equ:boundB} one can for example infer that the upper bound for the minimal random walk length will generally be higher when wanting to effectuate highly skewed scale-free networks with exponents close to two.

\Section{Protocol Definition}
\label{sec:Protocol}

The arguments laid out in the previous section suggest a rewiring protocol that consists of the following three basic operations: (1) In periodic intervals, a node $a$ selects an edge to a random neighbor $b$ that has not yet been rewired. (2) A random walk with the bias presented in equation \ref{equ:bias} is started to sample two nodes $x$ and $y$ with probabilities proportional to $\pi_x$ and $\pi_y$ respectively. (3) The edge $(v,w)$ is replaced by the edge $(x,y)$ and the latter is marked as having resulted from a rewiring operation. After all $m$ edges of the overlay have been rewired, a scale-free overlay is obtained whose exponent depends on the particular choice of the random walk bias defined in equation \ref{equ:bias}. In the algorithms \ref{alg:start} - \ref{alg:disconnect}, we give a detailed algorithmic description of the protocol. In these algorithms, $d_v$ denotes the degree of node $v$, $i_v$ is the ID of node $v$ and \emph{self} denotes the node at which the code is executed. We further assume that nodes have information about the IDs and the degrees of their nearest neighbors. 

The detailed algorithm of the main program loop that is responsible for initiating random walks is shown in algorithm \ref{alg:start}. Rewiring operations are initiated by nodes in regular intervals only for those edges that have not yet been rewired. By this means, at most $m$ rewirings are performed where $m$ is the number of edges in the initial random network topology. The number of rewiring operations and thus message transfers taking place within a certain time interval can be adjusted by choosing an appropriate (network-size dependent) \emph{delay} value. When a node with an unmarked edge wakes up, a rewiring operation is initiated. In order to prevent both endpoints of an edge to initiate rewiring operations for the same edge, rewirings are only started by the node with higher degree or - if the degrees are equal - by the node with the smaller ID. As we shall see later in section \ref{sec:Evaluation}, the choice of letting a rewiring be initiated by the better connected endpoint can improve the performance of the scheme. To find the endpoints of a new edge by which the previously unmarked edge shall be replaced, a node initiates a biased random walk through the overlay (lines $6-11$). In order to retain connectedness and prevent nodes from being isolated we further assume that only edges from nodes with degree greater than $1$ are rewired.

\begin{algorithm}[!ht]
\footnotesize
\begin{algorithmic}[1]
\caption{ Main Loop}
\label{alg:start}
\LOOP
\STATE $Sleep(delay)$
\IF{$neighbors.Count>marked.Count$}
\STATE $n = RandomUnmarkedNeighbor()$
\IF{$d_n>1\text{ \&\& } d_{self}>1$ \&\& $(d_{self}>d_{n}$ || $(d_{self}=d_{n}$ \&\& $i_{self}<i_{n}))$}
\STATE \COMMENT{Initiate random walk}
\STATE $msg.Hops \gets 0$
\STATE $msg.a \gets self$
\STATE $msg.b \gets n$
\STATE $msg.target \gets null$
\STATE $Send(\{walk, msg \}, n)$
\ENDIF
\ENDIF
\ENDLOOP
\end{algorithmic}
\end{algorithm}

When a node $v$ receives a random walk message, it needs to ensure that the message is forwarded with the bias given in equation \ref{equ:bias}. In algorithm \ref{alg:randomwalk}, this is done in lines $14-21$. Comparing the algorithm with the stochastic matrix $P$ defined in equation \ref{equ:bias}, here we select a neighbor uniformly at random and draw a random value uniformly in the interval $[0,1]$ that indicates whether the random walk transitions along this edge or whether it stays in the current node. One can imagine different schemes by which the two endpoints $v$ and $w$ of the new edge $(v,w)$ are sampled. The node initiating the rewiring could for example start one random walk for each endpoint of the new edge, collect the target nodes of both walks and connect them to each other. In order to simplify the implementation, in algorithms \ref{alg:start} and \ref{alg:randomwalk} we propose to sample both endpoints of the new edge in a single random walk of length $2 l$, assuming that after $l$ steps, the node at which the random walk currently resides is stored in the field $target$ of the message being forwarded. By this means, all information related to a rewiring operation is stored in the random walk message. Hence the node at which the random walk arrives after $2l$ steps has all information necessary to initiate the rewiring operation. For this, it creates a connection to the $target$ node stored in the message while initiating the deletion of the edge between node $a$ that has started the random walk and its neighbor $b$. As can be seen in algorithm \ref{alg:disconnect} a disconnection requires - apart from removing the edge - no further action at the side of the node from which the edge is removed. As shown in algorithm \ref{alg:connect}, both endpoints of the newly created edge mutually mark each other in order to prevent it from being rewired again in future invocations of the protocol. We emphasize that this is to prevent unnecessary rewiring operations and thus message exchanges rather than being required for the functioning of the protocol.

\begin{algorithm}[!ht]
\footnotesize
\begin{algorithmic}[1]
\caption{Node receives $\{walk, msg\}$}
\label{alg:randomwalk}
\STATE $msg.Hops \gets msg.Hops+1$	
\IF{$msg.Hops=l$}
	\STATE \COMMENT{Store Endpoint}
	\STATE $msg.target \gets self$
\ELSIF{$msg.Hops=2l$}
	\STATE \COMMENT{Rewire}
	\IF{$ !neighbors.Contains(msg.target)$ \&\&  $msg.target \neq self$}		
		\STATE $Send(\{disconnect, msg.a\}, msg.b)$
		\STATE $Send(\{disconnect, msg.b\}, msg.a)$
		\STATE $Send(\{connect, self\}, msg.target)$
		\STATE $Send(\{connect, msg.target\}, self)$
	\ENDIF
\ELSE
	\STATE $n \gets self.RandomNeighbor$
	\IF{$random.Next() \leq \frac{d_{self}}{d_n}\left(\frac{i_{self}}{i_{n}}\right)^{\frac{1}{\gamma_a-1}}$}		
		\STATE \COMMENT{Forward Random Walk}
		\STATE $Send(\{ walk, msg\}, n)$
	\ELSE
		\STATE \COMMENT{Self-Loop}
		\STATE $Send(\{ walk, msg \}, self)$
	\ENDIF
\ENDIF
\end{algorithmic}
\end{algorithm}

\begin{algorithm}[!ht]
\footnotesize
\begin{algorithmic}[1]
\caption{ Node receives $\{ connect, y\}$}
\label{alg:connect}
\STATE $neighbors.Add(y)$
\STATE $marked.Add(y)$
\end{algorithmic}
\end{algorithm}

\begin{algorithm}[!ht]
\begin{algorithmic}[1]
\footnotesize
\caption{ Node receives $\{ disconnect, b\}$}
\label{alg:disconnect}
\STATE $neighbors.Remove(b)$
\end{algorithmic}
\end{algorithm}

Concluding the description of the proposed protocol, we consider the size and number of messages that need to be sent across the network. Sampling the two endpoints of the new edge requires at most $2l$ messages \footnote{\emph{At most} $2l$ since self-loops are allowed to ensure aperiodicity of the underlying Markov chain. While a self-loop is a \emph{hop} of the random walk, it does not entail a message exchange.}, where $l$ is the number of steps taken by a single random walk to sample a node with a probability sufficiently close to the stationary distribution $\pi$. Once both endpoints of the new edge have been found, the rewiring requires two messages to disconnect nodes $a$ and $b$ and one message to connect to the node $target$ that has been stored in the random walk message.

Since the IDs of the initial node, its neighbors and the intermediate \emph{target}, as well as the current hop count need to be stored in the random walk message, the required number of bits for a message is logarithmic in the number $n$ of nodes in the system. Thus, the number of bits that need to be transferred per rewiring operation is $O(l\cdot log(n))$. Since exactly one rewiring operation is executed for each of the $m$ edges in the overlay topology, the total number of bits that need to be transferred in order to create a scale-free topology with the desired exponent is $O(m \cdot l \cdot log(n))$. We further require to store one additional bit per edge, indicating whether an edge has previously been rewired or not.

\Section{Evaluation}
\label{sec:Evaluation} 

Having given a description of the rewiring protocol as well as analytical arguments about its convergence behavior, in this section we present simulation results that have been obtained using an implementation of the proposed scheme. This evaluation is split up in two parts. In a first step, we seek to establish by simulation a practicable lower bound for the minimally random walk length $l$. We further study the influence of the initiating node's degree on the convergence time of a random walk. Based on these results, in a second step we then simulate the rewiring protocol and study its influence on a network's degree distribution.

\SubSection{Minimum Random Walk Length}
\label{sec:Evaluation:RandomWalk}

While theoretic asymptotic upper bounds for the required number of steps $l$ of the random walk have been presented in section \ref{sec:Bound}, here we empirically study the convergence behavior for a number of random walk lengths. By this we intend to derive a practicable random walk length that represents a reasonable trade-off between the imposed number of messages and the resulting total variation distance. We further intend to investigate how the minimum random walk length changes as the network sizes is varied. The following results have been obtained as follows. In each simulation run a number $R$ of random walks was started from a randomly chosen node in a random network topology. For each simulation of a random walk of length $l$, a hit counter was increased in the node at which the random walk resided in the $l$-th step. When $R$ random walks had been simulated, the total variation distance was computed based on the observed hit frequencies and the stationary distribution expected for the chosen random walk bias. Depending on network size and the minimum probability $\pi_{min}$ of the stationary distribution, the number of random walk iterations $R$ was chosen in a range between $10^6$ and $10^8$. In particular, it was chosen such that nodes with minimum stationary probability $\pi_{min}$ were expected to be hit reasonably often to argue about the total variation distance. The above procedure was then repeated for different random network realizations and starting nodes. Finally the minimum, maximum and average total variation distance of a simulation run was computed and the procedure was repeated for different network sizes, random walk biases and random walk lengths.

\begin{figure}[ht]
	\centering
		\includegraphics[width=6cm]{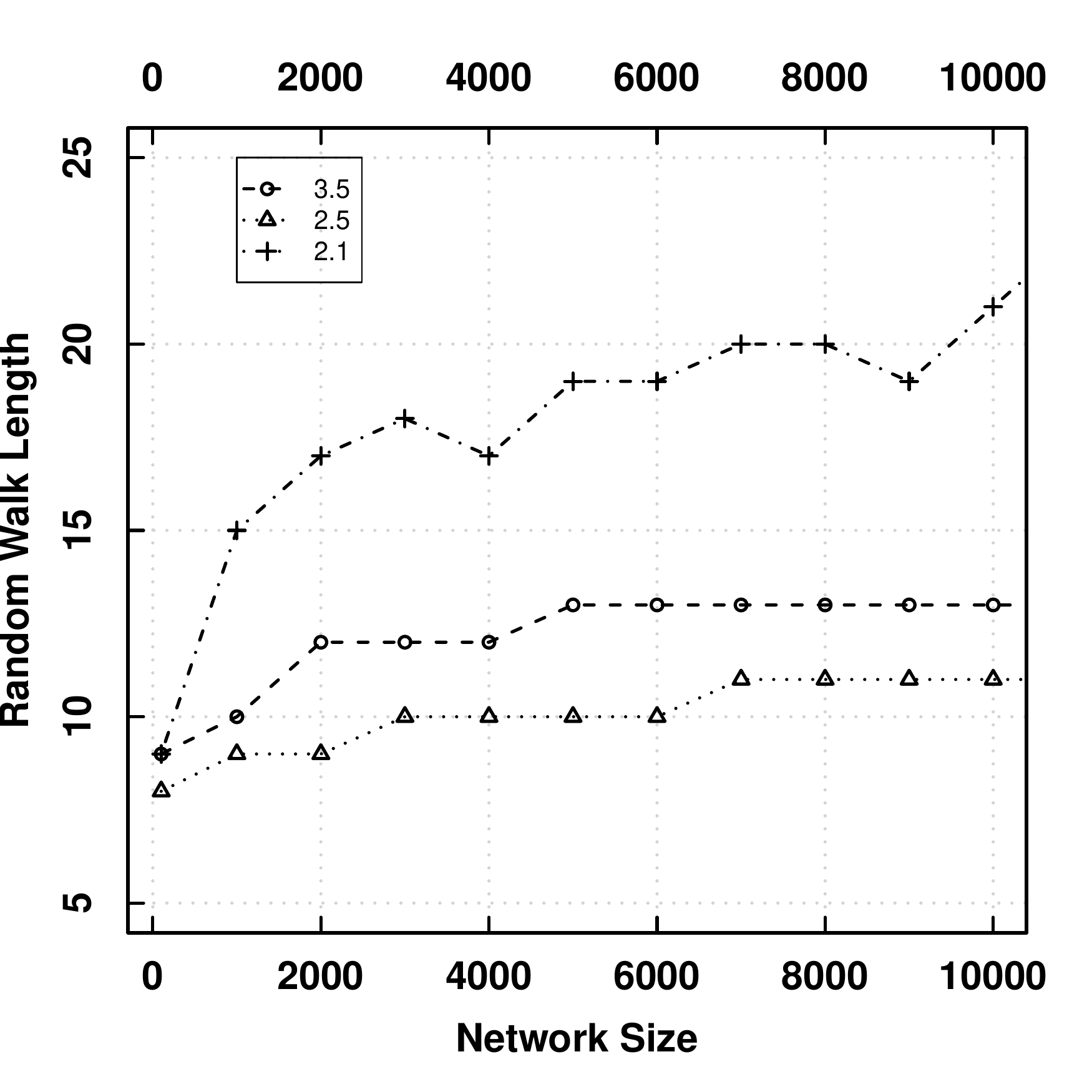}
		\caption{\label{fig:minLength} Minimum random walk length $l$ required to achieve $D(\pi(l), \pi) \leq 0.05$ in Barabási-Albert networks with random walk biases configured to effectuate exponents $2.1$, $2.5$ and $3.5$ (Lines are drawn to guide the eye)}
\end{figure}

Figure \ref{fig:minLength} shows the random walk length $l$ minimally required for the average total variation distance to fall below $\epsilon=0.05$ in scale-free Barabási-Albert (BA) networks randomly generated by the preferential attachment scheme presented in \cite{Barabasi1999}. Results are shown for different network sizes and for random walks configured to effectuate three different exponents $2.1$, $2.5$ and $3.5$. Rather than the linear scaling behavior suggested by the theoretical upper bound presented in section \ref{sec:Bound}, the observed required length $l$ rather scales in a sub-linear fashion. The observation that the actual convergence behavior is significantly better than the theoretical bound that can be obtained by a canonical path approach is consistent with the observations presented in \cite{Zhang2008} and indicates that the rewiring scheme can be implemented efficiently in practice. Further simulation results that have been obtained for Erdös/Rényi (ER) random graphs indicate that the number of steps required to achieve a total variation distance smaller than $0.05$ are in the same range as those for random power law graphs. Informally, this observed fast convergence can be attributed to the good expansion properties and the small diameter of both classical random graphs and random scale-free networks.

\begin{figure}[ht]
	\centering
		\includegraphics[width=6cm]{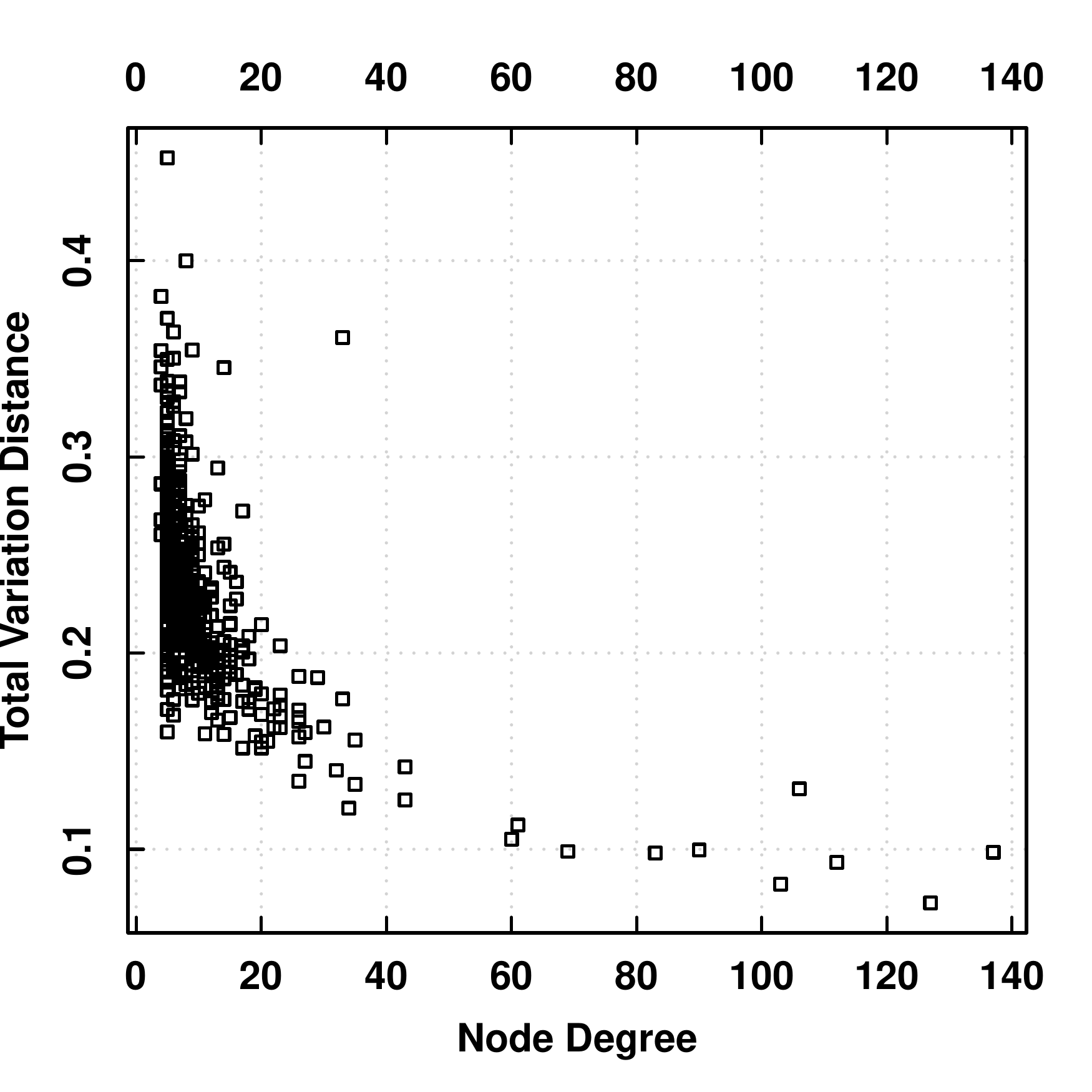}
		\caption{\label{fig:Correlation} Correlation between degree of starting node and average $D(\pi(l), \pi)$ in 1000 node Barabási-Albert networks with $\gamma=3$ and $l=5$}
\end{figure}

In networks with highly heterogeneous connectivity, a further interesting question is how the choice of the starting node of a random walk influences the total variation distance that can be achieved by a fixed random walk length. To investigate this, a number of random scale-free BA networks was created and a large number of random walks was started from each node of the network\footnote{Here \emph{large} again means sufficiently large to reasonably compute the total variation distance.}. The frequency with which nodes were target of random walks was recorded and the resulting total variation distance to the expected stationary distribution was computed for each starting node individually. Figure \ref{fig:Correlation} shows the relation between the degree of initial nodes and the total variation distance that was achieved in a representative simulation in random Barabási-Albert networks with $1000$ nodes, $l=5$ and a random walk bias to effectuate $\gamma=3$. Results suggest that random walks provide on average better convergence behavior when being started in highly connected nodes. In fact this is a rather intuitive result since a random walk starting at a high degree node can potentially reach a large number of nodes even in a single step. In the protocol presented in \ref{sec:Protocol}, this observation justifies the choice that rewiring operations for an edge $(i,j)$ are initiated by the node with higher degree.

\SubSection{Degree Distribution}

We now turn to the question how the rewiring protocol described in section \ref{sec:Protocol} influences the degree distribution of a network topology. All results presented in the following figures have been obtained for networks consisting of $5000$ nodes and roughly $25000$ edges. Initial topologies upon which the protocol was started were created using the Barabási-Albert preferential attachment model as well as the Erdös/Rényi model for classical random graphs. For the following measurements we used a random walk length that was long enough to achieve an average total variation distance smaller than $\epsilon=0.05$. Based on the results presented in the previous section a random walk length $l=20$ was chosen.

In each simulation run, the protocol presented in algorithm \ref{sec:Protocol} was applied by all nodes until all edges were rewired. The \emph{delay} interval between individual rewiring iterations was chosen such that- on average - a single rewiring took place per time unit. Thus, for the chosen network size and rewiring intensity, an adaptation cycle is expected to be completed after roughly $25000$ units of simulated time. The degree distribution of the network topology was computed each $200$ time units and a fit to the current degree distribution exponent was performed. For this, an $R$ implementation of the maximum likelihood power law fit procedure described in \cite{Clauset2007} was used. This procedure yields the fitted degree distribution exponent $\gamma_f$ that holds with maximum likelihood, the minimum network degree $d_{min}$ above which the fit holds, as well as the Kolmogorov-Smirnov (KS) statistic $D$. In general, better fits result in smaller values of $D$, thus allowing to evaluate whether the ``power law nature'' of the degree distribution is strengthened or fades away under the application of the rewiring scheme. All results are averages of at least $5$ independent applications of our protocol on randomly chosen network realizations of identical size. Simulation code, data analysis scripts, datasets, simulation videos as well some further graphical representations of results that could not be included in this article are available on the author's website.

\begin{figure*}[tb]
	\centering
		\subfigure[Average fitted exponent $\gamma_f$]{\label{fig:BA:Gamma} \includegraphics[width=5.4cm]{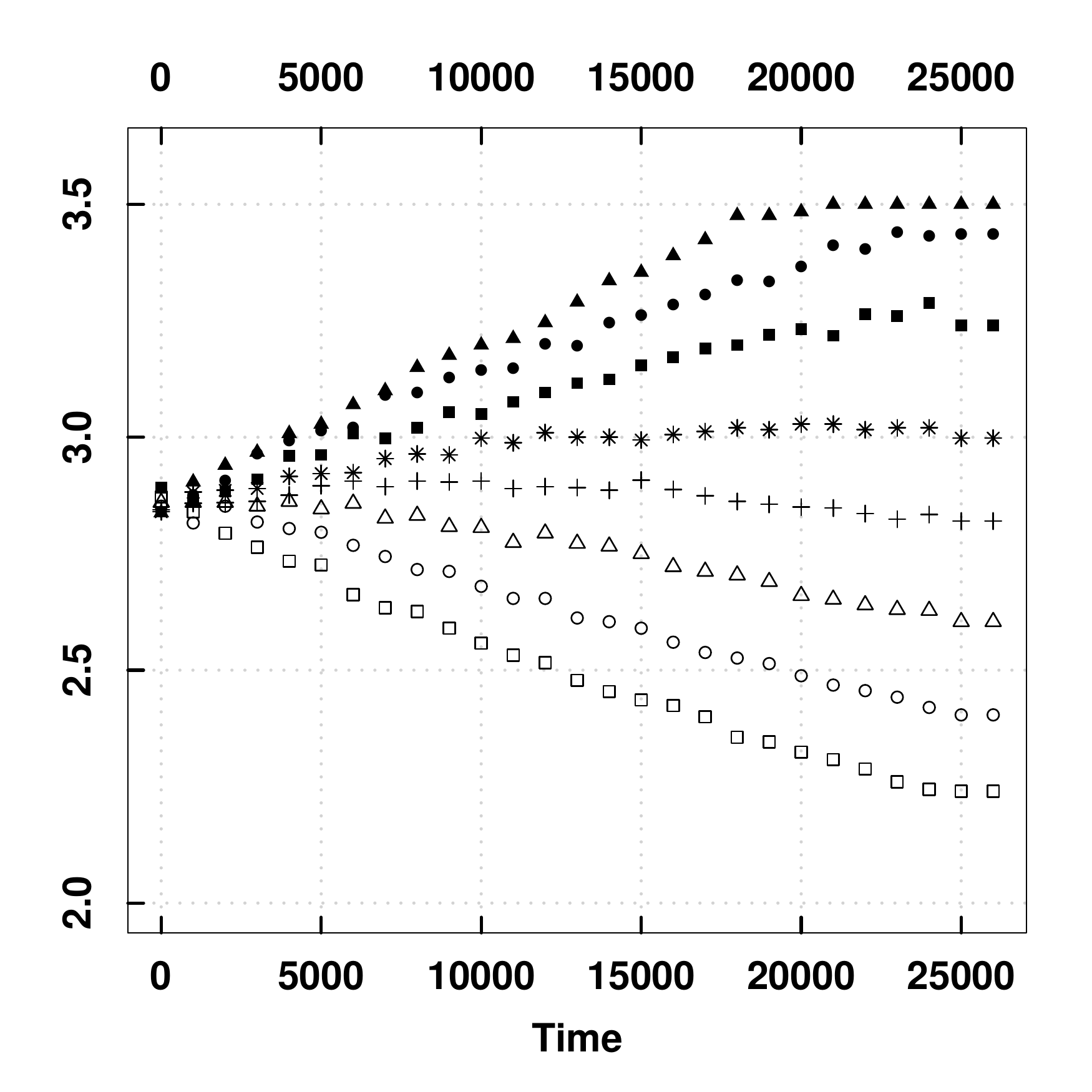}}
		\subfigure[Average Kolmogorov-Smirnov statistic $D$]{\label{fig:BA:D} \includegraphics[width=5.4cm]{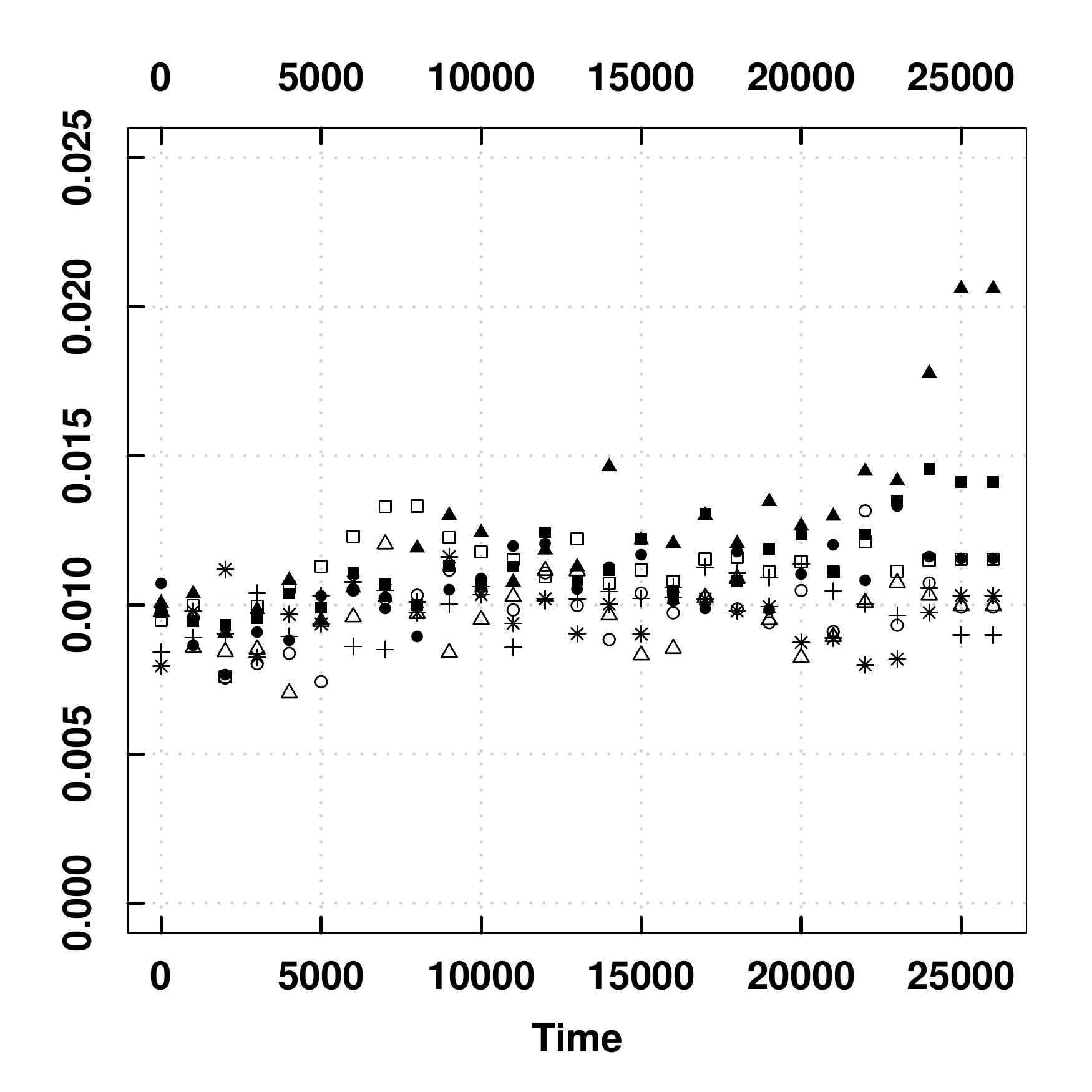}}
		\subfigure[Average maximum degree]{\label{fig:BA:MaxDeg} \includegraphics[width=5.4cm]{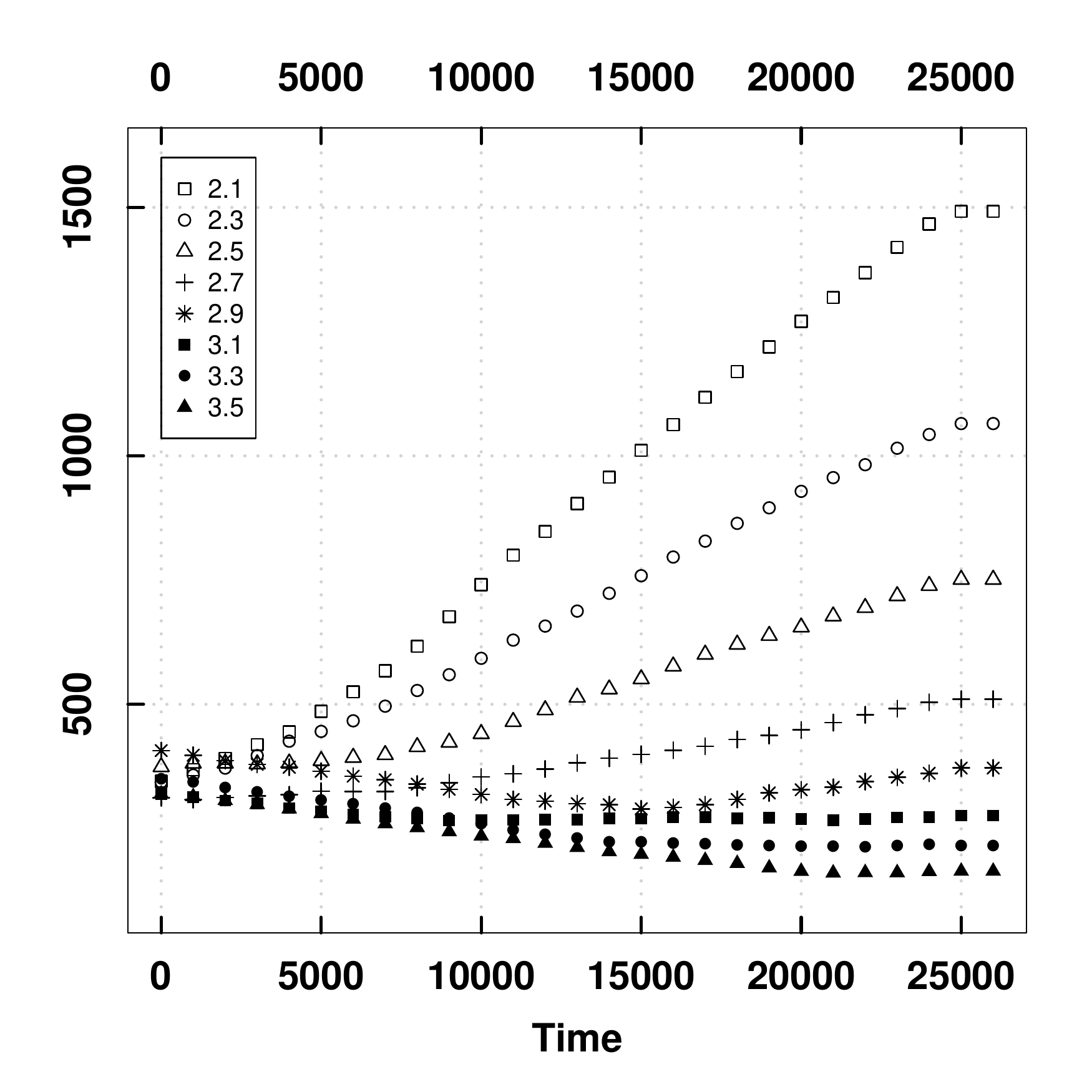}}  \qquad
		\subfigure[Average fitted exponent $\gamma_f$]{\label{fig:ER:Gamma} \includegraphics[width=5.4cm]{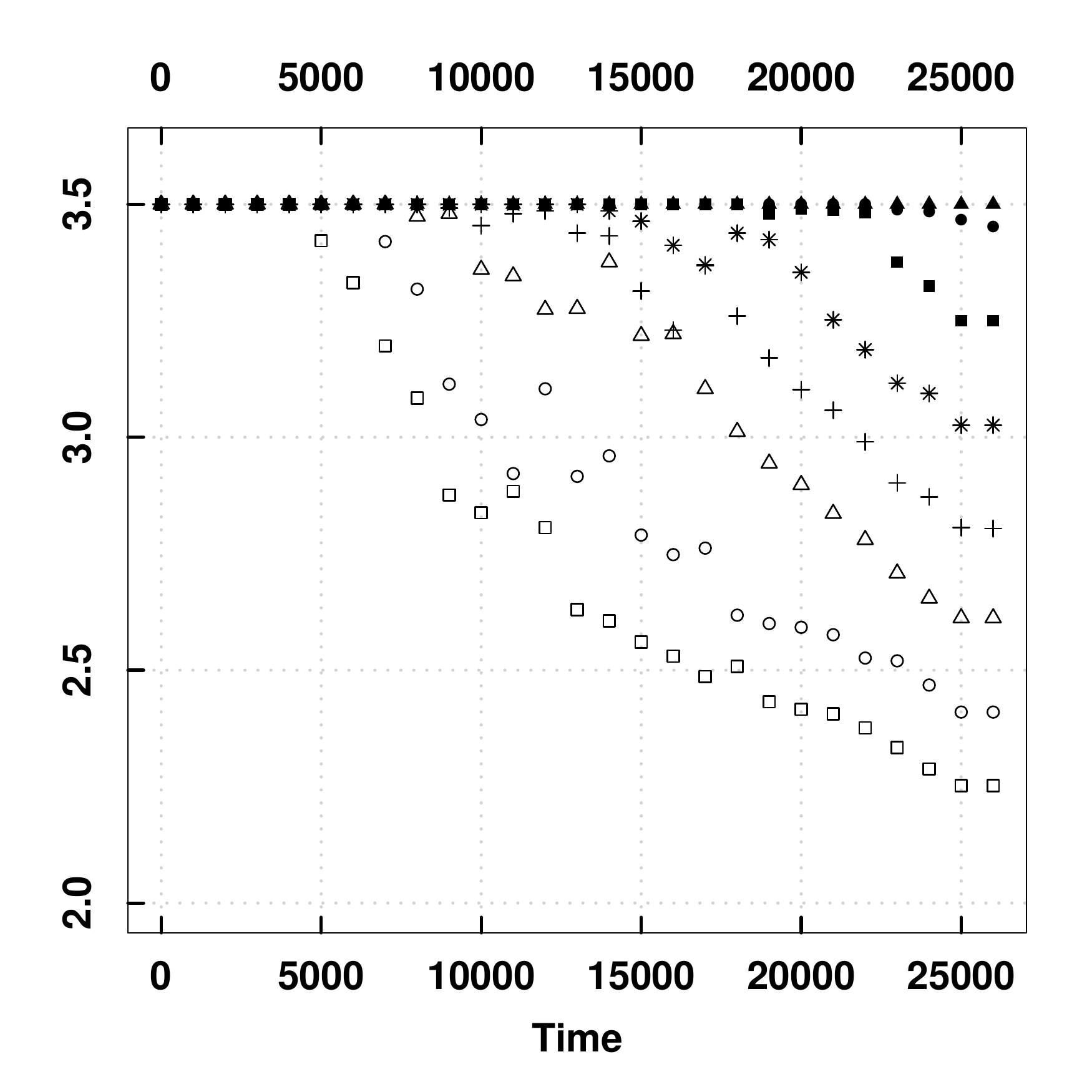}}		
		\subfigure[Average Kolmogorov-Smirnov statistic $D$]{\label{fig:ER:D} \includegraphics[width=5.4cm]{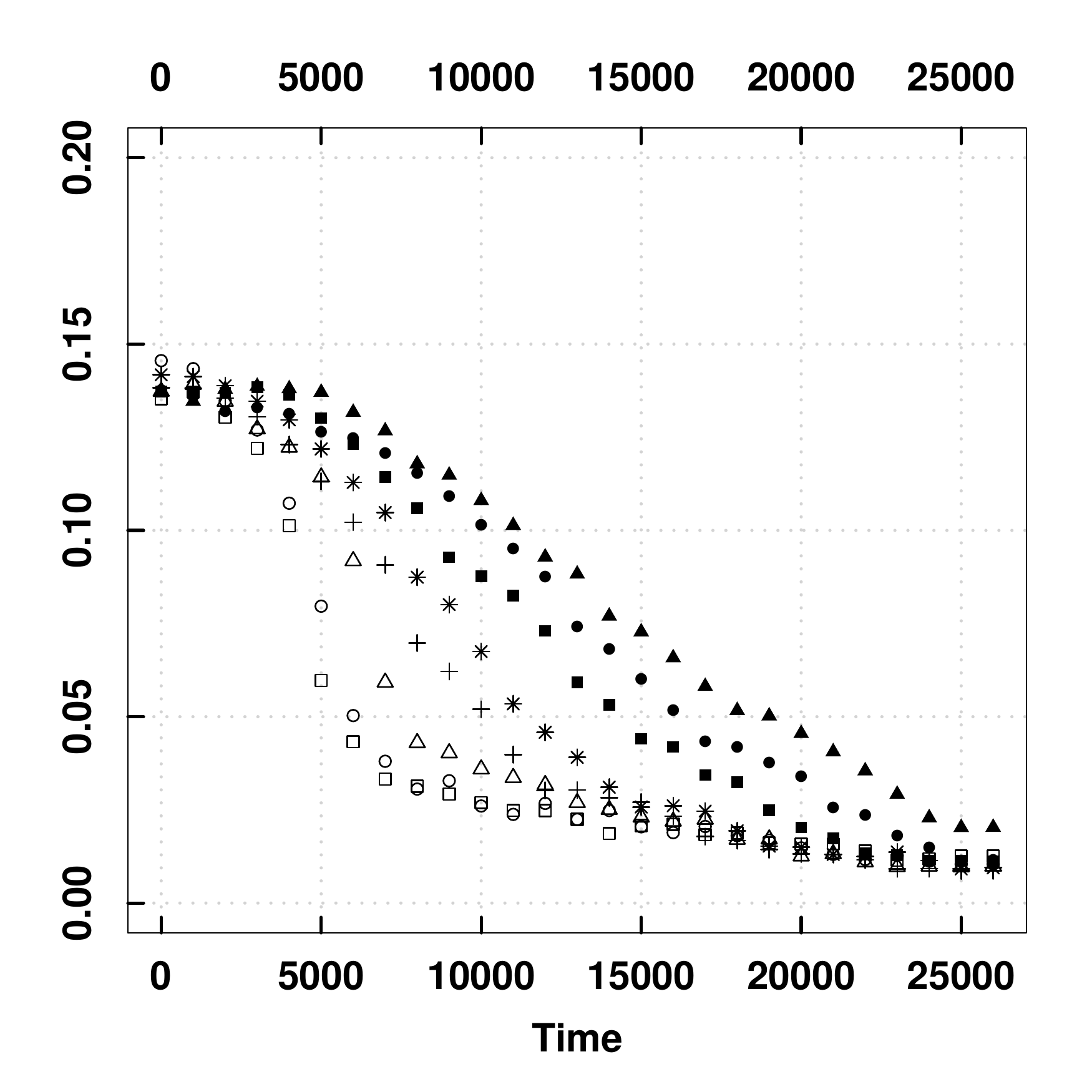}}
		\subfigure[Average maximum degree]{\label{fig:ER:MaxDeg} \includegraphics[width=5.4cm]{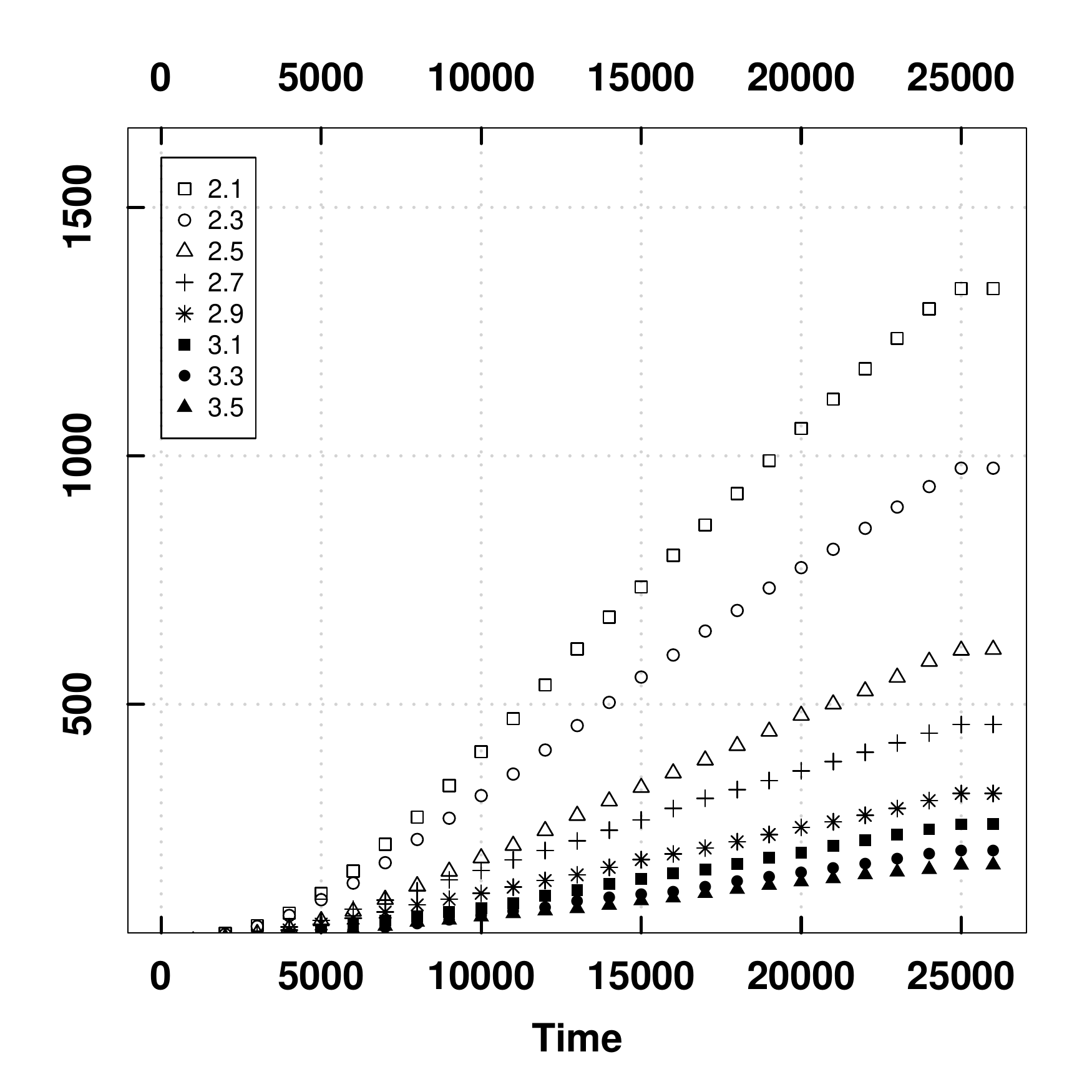}} \qquad		
	\caption{\label{fig:BA} Time Evolution of 5000 node Barab\'asi/Albert (a-c) and Erdös/Rényi (d-f) networks during adaptation runs with $\gamma \in [2.1, 3.5]$ and $l=20$}
\end{figure*}

Figures \ref{fig:BA:Gamma} and \ref{fig:ER:Gamma} show the effect of the proposed protocol on the degree distribution of a network that was initially created by the Barabási-Albert (BA), respectively Erdös/Rényi (ER) model. For BA networks, the average fitted exponent of the initial topology was on average $2.9$, while for ER networks the used fitting procedure yielded $3.5$ with an at least $10$-fold value of the KS-statistic $D$. The results suggest that the protocol does lead to an adaptation of the degree distribution exponent of the overlay topology. In particular, the evolution of the Kolmogorov-Smirnov statistic $D$ that is shown in Figure \ref{fig:BA:D} demonstrate that the scale-free characteristic of BA networks is preserved. The increase of the minimum degree above which the fit holds can be explained by the exponent-dependent finite-size effects in scale-free networks. For Erdös/Rényi networks, the roughly $10$-fold decrease of the KD-statistic $D$ that can be seen in \ref{fig:ER:D} indicates the emergence of scale-free characteristics, that is the power law fit to the degree distribution becomes more reliable. In Figures \ref{fig:BA:MaxDeg} and \ref{fig:ER:MaxDeg}, the evolution of the average maximum degree is shown. The results are consistent with the maximum degree expected in networks of the given size and with different degree distribution exponents. In Figure \ref{fig:Table}, the average fit parameters for the network topology eventually reached after adaptation are shown. The results demonstrate that - as expected from the underlying equilibrium model -  the protocol can be applied to transform arbitrary initial topologies into scale-free networks whose degree distribution is described by a power law with an exponent reasonably close to the intended value.

\begin{figure*}[tb]
	\centering
		\subfigure[Average fitted exponent $\gamma_f$]{\label{fig:BADynamic:Gamma} \includegraphics[width=5.4cm]{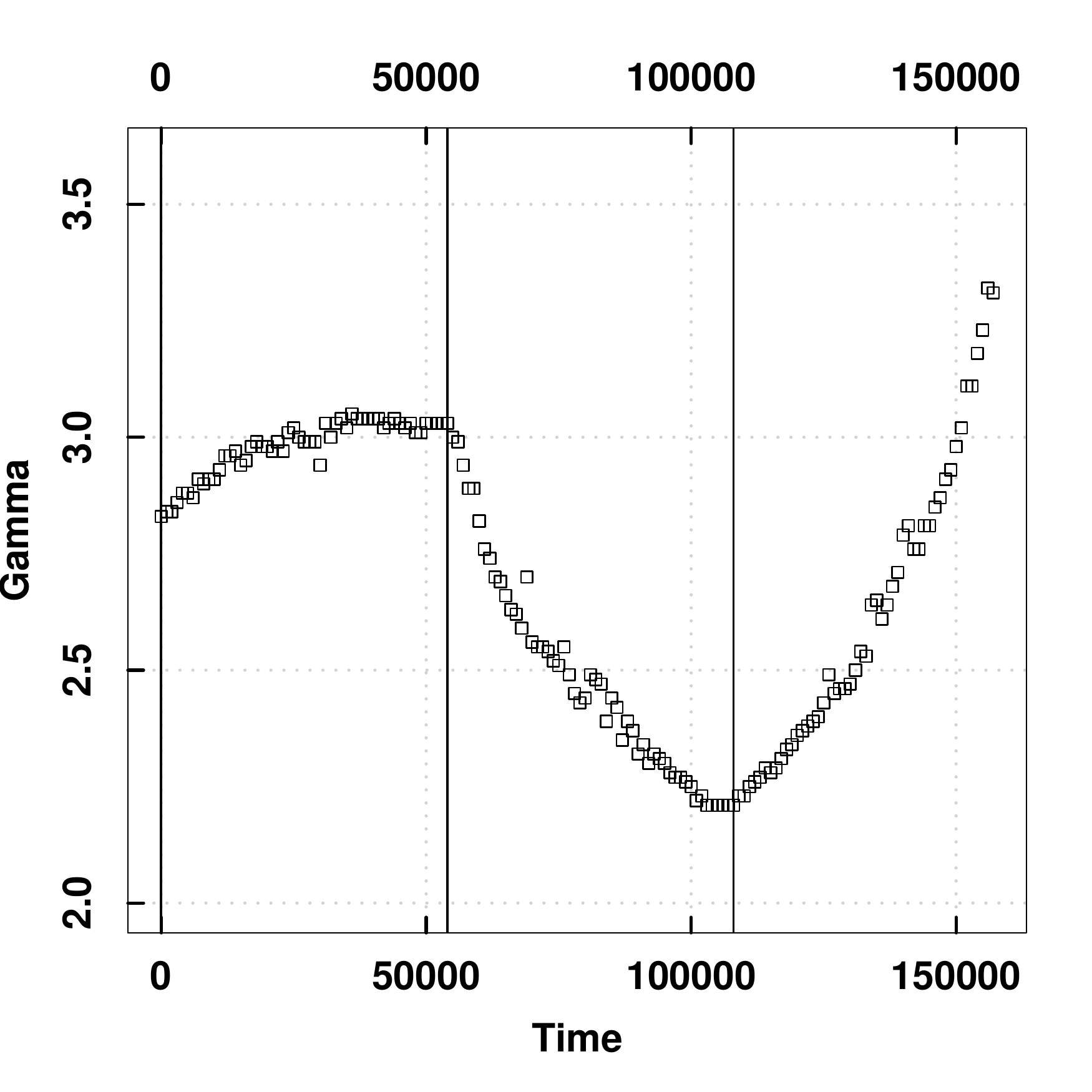}} 
		\subfigure[Average Kolmogorov-Smirnov statistic $D$]{\label{fig:BADynamic:D} \includegraphics[width=5.4cm]{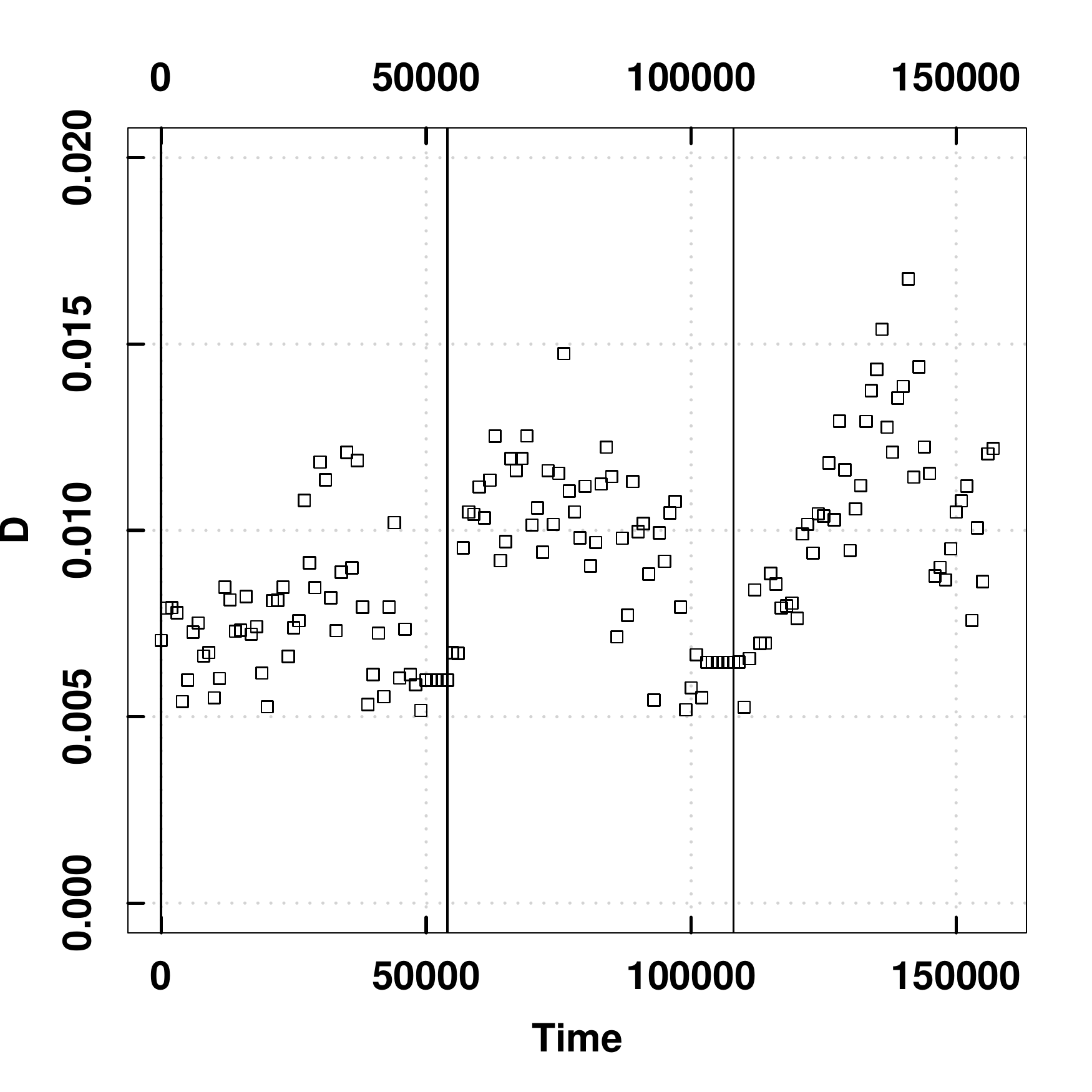}}
		\subfigure[Average maximum degree]{\label{fig:BADynamic:MaxDeg} \includegraphics[width=5.8cm]{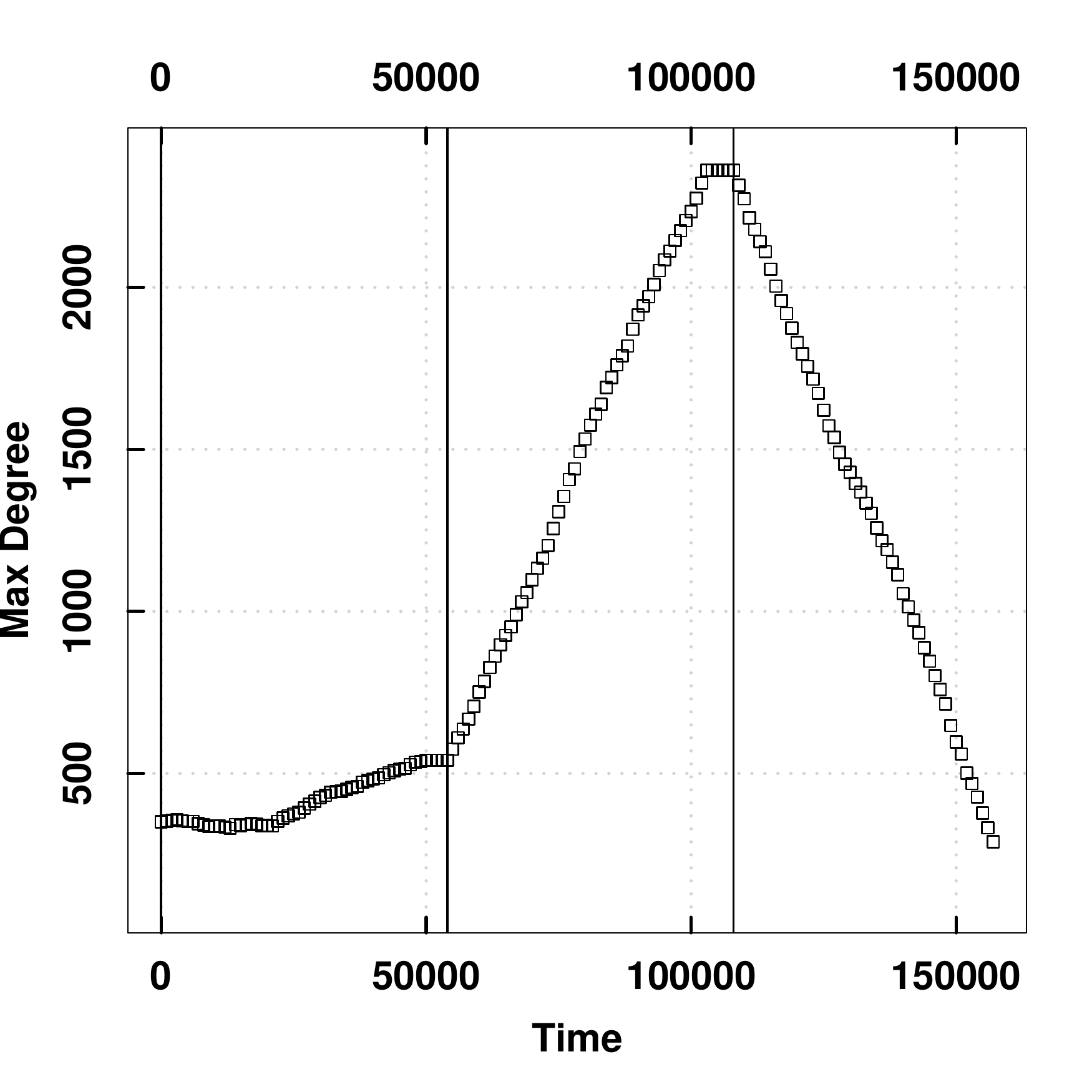}}
	\caption{\label{fig:Dynamic}Time Evolution of 10000 node Barabási/Albert network during multiple adaptation cycles with $\gamma_0=2.9$, $\gamma_{54000}=2.1$ and $\gamma_{108000}=3.5$. Start times of adaptation cycles are indicated by vertical lines.}
\end{figure*}

So far, we have only studied simulations using a single ``cycle'' of the proposed adaption protocol. In Figure \ref{fig:Dynamic}, results are shown for a simulation in which three adaptation cycles targeting different exponents were subsequently initiated in a Barabási/Albert network with $10^4$ nodes and roughly $5 \cdot 10^4$ edges. The chosen random walk length of $l=22$ was again consistent with the values found in section \ref{sec:Evaluation:RandomWalk}. In Figure \ref{fig:BADynamic:Gamma}-\ref{fig:BADynamic:MaxDeg}, time steps in which adaption cycles were started are indicated by vertical lines. The targeted degree distribution exponents were $2.9$, $2.1$ and $3.5$ respectively. Again the results indicate that the proposed scheme achieves the desired adaptation. Furthermore, Figure \ref{fig:BADynamic:D} shows how the ``power law nature'' of the degree distribution - and thus the scale-free characteristic of the network - temporarily fades during the adaptation while being restored near the ends of adaptation cycles.

\begin{figure}[!h]
	\centering		
	\scriptsize
	\scalebox{1.1}{
	\begin{tabular}{|c|c|c|c|c|c|c|c|c|c|c|}
			\hline
			& $\gamma_t$						& 2.1  		& 2.3 		& 2.5 	& 2.7 	& 2.9 	& 3.1 		& 3.3 		& 3.5  	\\
			\hline
			& $\gamma_f$						& 2.24 		& 2.40 		& 2.60	& 2.82	& 2.99	& 3.24 		& 3.44		& 3.5 	\\
			BA & $D$								& 0.012 	& 0.01		& 0.01	& 0.01	& 0.01	& 0.01		& 0.012		& 0.02 	\\
			& $d_{min}$								&	6.6			& 8.4			& 8.6		& 10		& 10		& 11.6		& 13.2		& 11.6 	\\
			\hline
			& $\gamma_f$						& 2.252		& 2.41		& 2.61	& 2.80	& 3.03	& 3.25		& 3.45		& 3.5  \\
			ER &	$D$								& 0.013		& 0.01		& 0.01	& 0.009	& 0.009	& 0.01		&	0.012		& 0.02 \\
			& $d_{min}$								&	7.6			&	8.8			& 9.2		& 9.6		&	11.4	& 12.6		& 12.2		& 12.4 \\
			\hline
		\end{tabular} }
		\caption{\label{fig:Table}Average fit parameters after adaptation with targeted exponents $\gamma_t \in [2.1,3.5]$ for $5000$ node Erdös/Rényi (ER) and Barabási/Albert (BA) networks with roughly $25000$ edges }		
\end{figure}

 \Section{Related Work}
 \label{sec:Related}
 
 During the last couple of years, a number of distributed approaches to the construction, maintenance and adaptation of probabilistically structured overlay topologies have been proposed. Here we briefly summarize a selection of approaches that are related to the present article. The use of random walks for the sampling of random participants in P2P overlays with good expansion (and thus Markov convergence) properties has been proposed in \cite{Gkantsidis2006,Zhong2005}. In particular, in \cite{Zhong2005} the use of biased random walks for a non-uniform random sampling of Peers is studied and analytical arguments for their convergence behavior are given. As argued in section \ref{sec:RandomWalk}, a similar random walk strategy constitutes the foundation for the adaptation scheme presented in this article. We finally emphasize that different approaches to a random sampling in P2P networks have been proposed as well, like for example the gossip-based topology management scheme considered in \cite{Jelasity2005a}. To date, it is however unclear how such alternative sampling mechanisms could be used in our particular scenario.
 
 Considering the problem of creating and adapting overlay networks with scale-free characteristics in a distributed fashion, it has been argued e.g. in \cite{Krapivsky2000} that the degree distribution exponent of scale-free networks can be tuned by adjusting the connection preferences of joining nodes. While this constitutes the basis for an adaption of growing networks, it remains unclear how the existing theoretical models can be implemented efficiently in practice. Considering practical networked systems, in \cite{Guclu2008} distributed strategies for the creation of scale-free overlays with connectivity cutoffs based on capacity constraints have been considered. Since the adaption of the degree distribution exponent does also change the maximum degree in the network, the schemes presented in \cite{Guclu2008} - although different in nature and intention - can be viewed as being related to the scheme presented in the present article. Finally, the problem of adapting the degree distribution exponent in scale-free overlay networks has been considered in own previous work \cite{Scholtes2008a}. However, in contrast to the protocol presented in the present article, no analytical arguments for the functioning of the scheme considered in \cite{Scholtes2008a} as well as its precise effects on the degree distribution exponent could be given. As such, the protocol can be viewed as a companion scheme to the distributed power law monitoring mechanism presented in \cite{Scholtes2008a}.

\Section{Conclusion}
\label{sec:Conclusion}

In this article, a simple adaptation protocol has been presented that is based on a rewiring strategy and the sampling of random nodes by means of a biased random walk. The protocol can be used to effectuate randomize scale-free overlay networks with adjustable degree distribution exponent and thus a tunable heterogeneity in connectivity. Apart from adapting the degree distribution exponent in scale-free networks, it is further suitable to transform arbitrary connected topologies to scale-free networks given that the expansion properties of the initial topology provide sufficiently fast convergence of the random walk strategy. In Barabási-Albert and Erdös/Rényi networks, the random walk length required to provide sampling probabilities that are acceptably close to the stationary limit are found to be significantly smaller than theoretical upper bounds. Based on empirical findings, we argue that the proposed protocol is thus a practicable approach to adapt probabilistically structured overlays for large-scale P2P systems. The performance of the protocol benefits from the fact that rewiring operations are preferentially started by high degree nodes as well as the observed superior convergence behavior of short-length random walks being started in hub nodes. In a future iteration of the protocol, it thus seems to be reasonable to choose the length $l$ of each random walk individually based on the degree of the node initiating it. A further potential improvement is the use of two-stage random walks which - in a first stage - preferentially move to hub nodes, and then - in a second stage - switch their bias to sample nodes according to the desired stationary distribution.

We conclude this article, by summarizing the main threats to validity and open issues. An important aspect in any practical application of the proposed scheme is the fact that a sufficiently efficient implementation of the scheme requires to accept moderate total variation distances. While this allows to keep the message overhead in an acceptable range, it limits the randomness of the resulting network topology. While small total variation distances suggest that the resulting deviation of properties from those of truly random networks are rather moderate, a further investigation of these effects is an open issue. Furthermore, although we have argued that simulations are a reasonable approach to establish empirical bounds on the required random walk length, the range of network sizes and topologies considered so far is fairly limited. A study of further topologies must thus be considered future work. Finally - and probably most important - the present implementation assumes per-node weights that are either based on fixed IDs or values chosen uniformly at random. An important next step is to extend the proposed scheme in way that characteristics like bandwidth capacity, uptime, processing power etc. of individual nodes are considered. While we believe that this can be done in a distributed fashion, the necessary adjustments of the proposed protocol must be considered future work. Due to these limitations, in its current state the proposed protocol is merely a first step towards practical Peer-to-Peer systems with scale-free overlay topologies that can efficiently be adapted in a directed and distributed fashion.

\bibliographystyle{latex8}

\end{document}